\documentclass[lettersize,journal]{IEEEtran}
\ifCLASSINFOpdf
 
\else

\fi

\usepackage{epsf}
\usepackage{graphics,graphicx,color,multirow,subfigure}
\usepackage{cite}
\usepackage{multirow}
\usepackage{amsmath}
\usepackage{enumerate}
\usepackage{txfonts}
\usepackage{graphicx}
\usepackage{subfigure}
\usepackage{algorithm}
\usepackage{algorithmic}
\usepackage{tikz}
\usepackage{booktabs}
\usepackage{epsf}

\begin{document}

\title{Streaming phishing scam detection method on Ethereum}
\author{\IEEEauthorblockN{Wenjia Yu\IEEEauthorrefmark{1},
		Yijun Xia\IEEEauthorrefmark{1},  Jieli Liu\IEEEauthorrefmark{2}, Jiajing Wu\IEEEauthorrefmark{1}\IEEEauthorrefmark{4}\\
	}
	\IEEEauthorblockA{\IEEEauthorrefmark{1}School of Computer Science and Engineering, Sun Yat-sen University, Guangzhou 510006, China\\}
	\IEEEauthorblockA{\IEEEauthorrefmark{2}School of Software Engineering, Sun Yat-sen University, Zhuhai 519082, China\\}

	\IEEEauthorblockA{\IEEEauthorrefmark{4}Correponding Author: wujiajing@mail.sysu.edu.cn}
     \thanks{This manuscript has been accepted by ISCAS 2023.}
	}
\maketitle
\begin{abstract}

Phishing is a widespread scam activity on Ethereum, causing huge financial losses to victims. Most existing phishing scam detection methods abstract accounts on Ethereum as nodes and transactions as edges, then use manual statistics of static node features to obtain node embedding and finally identify phishing scams through classification models. However, these methods can not dynamically learn new Ethereum transactions. Since the phishing scams finished in a short time, a method that can detect phishing scams in real-time is needed. In this paper, we propose a streaming phishing scam detection method. To achieve streaming detection and capture the dynamic changes of Ethereum transactions, we first abstract transactions into edge features instead of node features, and then design a broadcast mechanism and a storage module, which integrate historical transaction information and neighbor transaction information to strengthen the node embedding. Finally, the node embedding can be learned from the storage module and the previous node embedding. Experimental results show that our method achieves decent performance on the Ethereum phishing scam detection task. 

\end{abstract}

\IEEEpeerreviewmaketitle

\section{INTRODUCTION}


Ethereum is currently the largest blockchain platform supporting smart contracts. It is facing a huge ecological security crisis with numerous scams \cite{wood2014ethereum}. Among them, phishing scams occupy a large proportion \cite{anita2019blockchain}. Therefore, how to detect phishing scams on Ethereum is increasingly attracting attention from researchers.


In general, existing methods for detecting Ethereum phishing scams typically use traditional feature engineering methods \cite{chen2020phishing} or graph representation learning methods \cite{WhoAreThePhisher, TEGDetector, li2021self} to identify phishing scam nodes. These methods model the Ethereum transaction network as a graph, with nodes representing Ethereum accounts and edges representing transactions between accounts. The former extracts the transaction features of the nodes and trains the classifier of the machine learning model to complete the detection task. The latter uses random walk or graph neural network to learn the node embedding. Phishing scam nodes are then identified by node classification. These methods all use manual-designed features that rely on expert knowledge and extract node transaction records over a certain time domain as node features (e.g., number of transactions, total transaction amount, number of degrees, etc.). In addition, most of these methods are static. They all model the Ethereum transaction network as a static graph. The extracted node features are also static, which is because these node features are statistical features over a period of time and are not real-time.

However, new transactions are generated on Ethereum every moment, and the Ethereum transaction network changes dynamically over time. Static methods do not capture the changes that occur in the network \cite{kim2018review}. Some methods take into account the dynamic nature of the Ethereum transaction network and therefore use a discrete-time slice method to model the Ethereum transaction network at different moments, where the change of the transaction network is observed as a collection of static network snapshots over time\cite{li2021self}. Although such methods consider the network changes, the choice of time slice length still relies on expert knowledge and is not extensible enough.


In general, there are still two remaining problems: ($1$) Lack of dynamicity. At the level of data pre-processing, the use of statistical features over time as node features undermines the real-time nature of the model. At the method level, static methods do not capture the dynamic nature of the network. Although discrete-time slice methods take into account the evolution of the transaction network over time, they capture temporal information at a very coarse level which leads to a loss of information between snapshots. ($2$) Primitive information abstraction. The account in real Ethereum transaction scenarios does not have primitive features. Most of the existing methods manually count the transaction features in a certain time domain to be used as node features, which are inefficient and non-automated.


To address the above problems, in this paper, we propose a streaming phishing scam detection method. First, we use transaction edge features instead of node features. The advantage of using edge features is that it avoids the manual step of counting node features, thus enabling the processing of each new transaction that occurs in real-time. After that, in order to perform representation learning on nodes, we design a storage module to store the node's historical transaction information and its neighbors' transaction information, and a broadcast module to propagate the node's transaction information to its neighbors. Both the historical transaction information of nodes and the neighbor information can enrich the representation of nodes.

Our main contributions to this work consist of the following:

\begin{itemize}
    \item Our methods can capture the evolution and the continuous time fine-grained temporal dynamics of the Ethereum transaction network, which can improve the accuracy of identifying phishing scam nodes. 
    
    \item The first work utilize edge features instead of manually static node features. It can enable real-time detection of each newly generated transaction for the purpose of streaming detection of phishing scam nodes. 

\end{itemize}

Section~\ref{sec:data} introduces the raw Ethereum data and the ground-truth labels including phishing and other kinds of non-phishing labels we collected. Section~\ref{sec:methods} introduces the overall detection framework. The comprehensive experiment results on the real Ethereum network will be given in Section~\ref{sec:experiment}. Finally, we will give a brief conclusion in Section~\ref{sec:conclusion}.

\section{DATA\label{sec:data}}

\subsection{Raw Data Collection}
Our raw data of Ethereum transactions are obtained from the website http://xblock.pro, one of the widely used blockchain data platforms in the academic community. We collect data from block height 8,000,000 to block height 8,999,999. The raw data for a specific transaction record includes eleven components: $(1)$ Sender account of a transaction. $(2)$ Recipient account of a transaction. $(3)$ Amount of ETH transferred. $(4)$ Maximum gas allowed to 
be consumed. $(5)$ The number of gas consumed. $(6)$ Price of one unit of gas in the transaction. $(7)$ A unique 66-character identifier of a transaction. $(8)$ Time when a transaction is mined. $(9)$ Block number. $(10)$ Whether the sender is the contract account. $(11)$ Whether the recipient is the contract account.  

\subsection{Labels}
We collect the phishing labels and the non-phishing labels of accounts from Etherscan (https://etherscan.io/), a famous block explorer and analytic platform for Ethereum, which website reports various cybercrimes and scam accounts on Ethereum. We crawl all the reports about phishing scams before September 20th, 2021, and label the accounts reported by Etherscan as phishing accounts. At last, we construct a set of ground-truth account labels including 426 phishing labels and 34,960 non-phishing labels within the one million blocks.

\section{METHODS\label{sec:methods}}

\begin{figure*}[t]
 	\centering
 	\includegraphics[width=\linewidth]{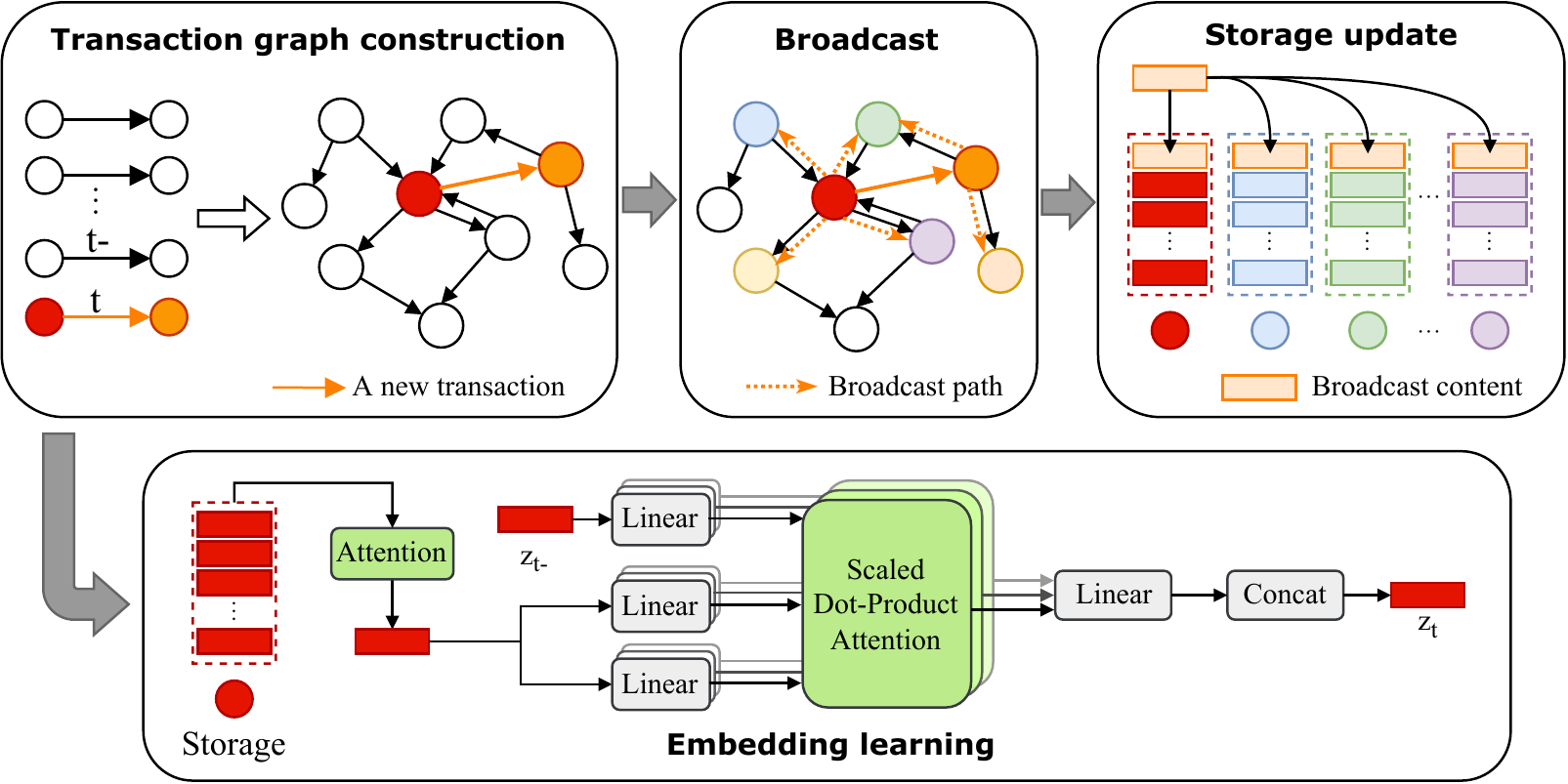}
 	\caption{The overall architecture of our streaming phishing scam detection method. When a new transaction occurs at time $t$, the target node (the red node here) will update its embedding. Through the embedding learning module, the embedding $z_t$ of the target node is learned from its storage at time $t-$. Meanwhile, the storage of the target node and the neighbors update after the broadcast module. }
 	\label{fig:framework}
\end{figure*}

Fig. \ref{fig:framework} shows an overview of our proposed detection method. It abstracts the transactions as a series of temporally ordered events and uses three modules to learn the representation of nodes. Finally, The model converts the detection task into a classification task to classify phishing scam nodes for detection purposes.

\subsection{Preliminaries}

In order to extract more comprehensive edge features, we consider the impact of smart contracts on transactions. Accounts on Ethereum are divided into external accounts and contract accounts\cite{zheng2018blockchain}. Smart contracts on Ethereum are created and called by external accounts, which represent contract accounts \cite{chen2020understanding}. If an external account calls multiple smart contracts when completing a transfer thereby generating multiple internal transaction records, we treat these transactions as one complete transaction for the purpose of extracting edge features. We extract the edge features in 16 dimensions, including Block number, current amount of ETH transferred, timestamp, the number of smart contracts called, etc. 

Formally, We model the Ethereum transaction network as a continuous-time dynamic graph(CTDG) for streaming detection. We view transactions as a series of time events $\delta(t) = (v_i, v_j, e_{ij}, t)$ order by timestamp, which means node $v_i$ initiates a transaction to node $v_j$ at time $t$. The edge feature matrix $e_{ij}\in \mathbb{R} ^{M\times d}$ consists of all temporal events in a CTDG, where $M$ is the number of events and $d$ is the dimension of the edge feature. Finally, Ethereum transaction network can be represented as $ \mathcal{G} = \{\delta(t_1), \delta(t_2),...\}$.


\subsection{Broadcast mechanism}
When a transaction is executed on Ethereum, the transaction is broadcasted to the entire Ethereum network. After receiving the broadcast, other nodes will check whether the transaction is valid or not. If the transaction is valid, a miner, a special node in the Ethereum, will pack this successful transaction into a block and add it to the blockchain \cite{holub2018coinhoarder}. 

Inspired by this, we also design a broadcast mechanism for our method. 
Like most graph convolutional networks(GCN) methods, our method is based on the assumption that the state of a node in a graph is always influenced by its surrounding nodes \cite{gcn}.
When a transaction event occurs, the broadcast mechanism encodes the event and broadcasts it to the neighboring nodes of the nodes involved in the transaction. The broadcast module in Fig. \ref{fig:framework} shows an example of the process of broadcasting. For a transaction that occurs at time $t$, we use the following mathematical formula to generate broadcast content:
\begin{equation}
    \ c(t) = z_{u}(t) + \xi_{uv} + z_{v}(t),
\end{equation}
where $\ c(t)$ is the broadcast content at time t, $z_{u}(t)$ denotes the embedding of the node $u$ at time $t$, $z_{v}(t)$ denotes the embedding of the node $v$ at time $t$, $\xi_{uv} $ denotes one of the edge features between node $u$ and node $v$ at time $t$. After generating $\ c(t)$, we broadcast $\ c(t)$ to the node $u$, $v$ and their first-order neighbours. 

Since the short active period of accounts on Ethereum, we can rely on the time-first principle and select the $k$ nodes that are closest to the target node transactions to broadcast.

\subsection{Storage module} When a node receives the broadcast contents from the other nodes, it stores the broadcast contents in its own storage module. To save the memory overhead of the algorithm, we set the storage module to a fixed-length first-in-first-out (FIFO) queue. The storage module will generate a vector $s(t)$ by aggregating the broadcast contents to support the model update the node embedding. 
It is obvious that the contents in the storage module are arranged in temporal order, and the earlier the content is stored, the less important it becomes. Based on this, we introduce a decay factor $\alpha = (\alpha_{1}, \alpha_{2}, ..., \alpha_{m})$ that helps aggregating contents in the storage module:
\begin{equation}
    s(t) = agg\left(\\c_{1}(t), \\c_{2}(t),...,\\c_{m}(t)\right) ,
\end{equation}
\begin{equation}
    agg\left(\\c_{1}(t), \\c_{2}(t),...,\\c_{m}(t)\right) = \frac{1}{m} \sum_{i=1}^m \alpha_{i}\cdotp \\c_{i}(t),
\end{equation}
where $m$ denotes the length of storage module, $\alpha_{i}$ is a hyperparameter set empirically, $c_{i}(t)$ denotes the $i$-th broadcast content stored in the storage module. 

%
\subsection{Embedding} Finally, the model will update the embedding $z_{i}(t)$ of node $i$ at time $t$, which can be further utilized in phishing scam detection. We use the embedding of the node $i$ just before time $t$, and the embedding of the storage module that belongs to node $i$ as the inputs of muti-head attention \cite{vaswani2017attention}, and learn the relationship of the input by muti-head attention to get the final embedding of node $i$. The hidden mechanism of a single-head attention layer can be defined as:
\begin{equation}
\begin{aligned}
    Attn(Q, K, V) &=softmax\left( \frac{QK^{T}}{\sqrt{d}} V\right), \\
    Q &= z(t-)W_{Q}, \\
    K & = s(t)W_{K},\\
    V & = s(t)W_{V},
\end{aligned}
\end{equation}
where Q, K and V are all vectors that denote `queries', `keys', and `values', respectively. The function $Attn(Q, K, V)$denotes the scaled dot-product attention \cite{vaswani2017attention} that takes a weighted sum of the entity $V$ where the weights are given by the interactions of entity $Q-K$ pairs. Essentially it is a mapping of a `query' to a `key'-`value' pair. $W_{Q},W_{K},W_{V}$ denote the projection weight matrices that are employed to learn the suitable $Q, K, V$ to create the performance attention output. Here we use the embedding of node at last update time $t-$ as the input $Q$, the embedding of the node`s storage module as the input $K, V$ to update node embedding at time $t$, which means the attention module can capture the relationship between node embedding at time $t-$ and the neighbors of node.

To avoid possible bias caused by single attention mechanism, we use a multi-head attention mechanism to learn a better embedding. Multi-head can form multiple subspaces and force model learning different aspects of information. Considering the dimensions of our features, we choose three-head attention:
\begin{equation}
\begin{aligned}
    head_{i} &= Attn(Q_{i}, V_{i}, K_{i}), i=1,2,3, \\
    MultiHead(Q,K,V) &= Concat(head_{1}, head_{2}, head_{3})W_{o},
\end{aligned}
\end{equation}
where $W_{o} \epsilon R^{d\times d} $, $head_{i}\epsilon R^{\frac{d}{3}}$. And then, the node embedding at time t are represented as:
\begin{equation}
\begin{aligned}
    \widetilde{z}(t) = MultiHead(Q,K,V).
\end{aligned}
\end{equation}

After that, we use a multi-layer perceptron to combine the node embedding at time $t$ with the node embedding at time $t-$. This corresponds to re-fusing the reference node representation with the aggregated information:
\begin{equation}
\begin{aligned}
    z(t) = MLP(z(t-) \Vert \widetilde{z}(t)).
\end{aligned}
\end{equation}

\section{EXPERIMENTS\label{sec:experiment}}

\subsection{Dataset}
After we obtain the Ethereum transaction records from block height 8,000,000 to block height 8,999,999, we extract the first-order neighborhood transaction information of the labeled accounts to validate the performance of our method. For all datasets, train/val/test sets are split with a ratio of 70/15/15.

\subsection{Baseline methods}
We use three types of baselines. $1)$ Feature engineering methods: Decision Tree classifier with two different ways of feature aggregation. $2)$ Static graph representation methods: GraphSAGE \cite{hamilton2017inductive}, Trans2Vec \cite{WhoAreThePhisher}. $3)$ Dynamic graph representation methods:DyRep \cite{trivedi2019dyrep}, JODIE \cite{van2010comparing}, TGN \cite{rossi2020temporal}. To be fair, we use the same data processing and splitting methods for all these methods. 

\subsection{Comparisons with baselines}
Table \ref{table:result} indicates the phishing scam detection experiment results of our methods and baselines. Since all static methods require primitive node features as input, in order to experiment with these static methods, we manually aggregate edge features into node features. It can be seen that the performance of the DT-Mean method and the DT-Sum method is average. GraphSAGE and Trans2Vec obviously perform worse than the two feature engineering methods. Trans2Vec performs the worst due to the fact that Trans2Vec initially used a dataset with a positive and negative sample distribution close to 1:1, while we used a dataset with a large gap between the positive and negative sample ratios in order to simulate the real data distribution as much as possible. Compared with feature engineering methods, static graph representation learning methods introduce additional static structure information of the graph, which adds noise and makes its performance inferior to that of feature engineering methods. 

Dynamic graph representation learning methods take into account the change of the graph structure over time, consider the dynamic structural information of the graph, and ultimately perform better than other methods. It can be attributed to the fact that dynamic methods capture information about the changes in the network over time, resulting in a more accurate representation of the nodes. Among them, DyRep performs well on the AUC metric and poorly on the TPR metric, suggesting that DyRep is not adapted to handle datasets with positive and negative sample imbalance. Both JODIE and TGN outperform the remaining two types of baselines overall but perform poorly on the most important TPR metric. In contrast, our method achieves good competitive performance on the AUC and TPR metrics compared to the other methods. The characteristics of Ethereum datasets that differ from traditional graph datasets make existing dynamic methods perform less well than ours on these datasets, due to the fact that our method is designed to fit the Ethereum transaction scenario by utilizing unlabeled nodes to enhance broadcast and storage capabilities of the model. In addition, our method directly utilizes edge features for detection purposes without relying on the primitive features of the nodes.


\subsection{Ablation study}
In this section, we study the influence of different modules in our method. As shown in Table \ref{table:ablation}, different modules bring different degrees of improvement to the detection performance in phishing scam detection. In particular, the storage module makes the highest contribution to performance. It indicates that in the Ethereum transaction network, the neighbor transaction information of the target node is crucial. We also find that encoding the importance of historical information in the storage module improves the performance. It suggests that the importance of transaction information varies across neighbors. Removing the broadcast mechanism means that our framework relies only on the historical transaction information of the target node to learn the node embedding, which does not lead to accurate detection results.
\begin{table}[]
\renewcommand{\arraystretch}{1.2}
\centering
\caption{AUC, TPR, FPR (\%)  for the phishing scam detection.}
\label{table:result}
\begin{tabular}{l|lll}
\hline
Method         & AUC            & TPR            & FPR           \\ \hline
DT-Mean        & 62.30          & 55.27          & 3.64          \\
DT-Sum         & 64.34          & 58.28          & 4.73          \\
GraphSAGE      & 51.25          & 57.09          & 6.25          \\
Trans2Vec      & 73.50          & 43.04          & 1.02         \\
DyRep          & 88.35          & 25.00          & 1.17          \\
JODIE          & 77.09          & 64.06          & \textbf{1.00} \\
TGN            & 89.77          & 73.44          & 1.84          \\ \hline
Ours           & \textbf{96.26} & \textbf{90.08} & 1.79          \\ \hline
\end{tabular}
\end{table}

\begin{table}[]
\renewcommand{\arraystretch}{1.2}
\centering
\caption{Ablation experiment on scam detection.}
\label{table:ablation}
\begin{tabular}{@{}llll@{}}
\toprule
Enabled Module                             & AUC   & TPR   & FPR   \\ \midrule
w/o decay factor & 94.35 & 65.31 & 1.23  \\
w/o broadcast mechanism                         & 78.34 & 32.81 & 1.00  \\
w/o storage module                              & 65.03 & 27.48 & \textbf{0.05} \\
Ours        & \textbf{96.26} & \textbf{90.08} & 1.79  \\ \bottomrule
\end{tabular}
\end{table}

\section{CONCLUSIONS\label{sec:conclusion}}
In this paper, we proposed a method for streaming phishing scam detection. The method utilized edge features to capture every transaction streamingly. When a new transaction emerges, it broadcasts the transaction information to its neighbors through a broadcast mechanism. At the same time, it updates the current node embedding representation using the neighboring transaction information in the node storage module and the node's embedding at the previous moment. It can achieve strong representations of nodes by combining neighborhood characteristics, transaction characteristics, and node history characteristics. Experiments indicated that our method could adapt to the real Ethereum transaction network and its performance on streaming phishing scam detection tasks outperforms state-of-the-art algorithms.


\bibliographystyle{IEEEtran}
\bibliography{ref.bib}

\begin{thebibliography}{10}
\providecommand{\url}[1]{#1}
\csname url@samestyle\endcsname
\providecommand{\newblock}{\relax}
\providecommand{\bibinfo}[2]{#2}
\providecommand{\BIBentrySTDinterwordspacing}{\spaceskip=0pt\relax}
\providecommand{\BIBentryALTinterwordstretchfactor}{4}
\providecommand{\BIBentryALTinterwordspacing}{\spaceskip=\fontdimen2\font plus
\BIBentryALTinterwordstretchfactor\fontdimen3\font minus
  \fontdimen4\font\relax}
\providecommand{\BIBforeignlanguage}[2]{{%
\expandafter\ifx\csname l@#1\endcsname\relax
\typeout{** WARNING: IEEEtran.bst: No hyphenation pattern has been}%
\typeout{** loaded for the language `#1'. Using the pattern for}%
\typeout{** the default language instead.}%
\else
\language=\csname l@#1\endcsname
\fi
#2}}
\providecommand{\BIBdecl}{\relax}
\BIBdecl

\bibitem{wood2014ethereum}
G.~Wood, ``Ethereum: A secure decentralised generalised transaction ledger,''
  \emph{Ethereum project yellow paper}, vol. 151, pp. 1--32, 2014.

\bibitem{anita2019blockchain}
N.~Anita and M.~Vijayalakshmi, ``Blockchain security attack: A brief survey,''
  in \emph{Proceedings of the 10th International Conference on Computing,
  Communication and Networking Technologies (ICCCNT)}.\hskip 1em plus 0.5em
  minus 0.4em\relax IEEE, 2019, pp. 1--6.

\bibitem{chen2020phishing}
W.~Chen, X.~Guo, Z.~Chen, Z.~Zheng, and Y.~Lu, ``Phishing scam detection on
  ethereum: Towards financial security for blockchain ecosystem.'' in
  \emph{Proceedings of the 17th International Joint Conferences on Artificial
  Intelligence}.\hskip 1em plus 0.5em minus 0.4em\relax IJCAI.org, 2020, pp.
  4506--4512.

\bibitem{WhoAreThePhisher}
J.~Wu, Q.~Yuan, D.~Lin, W.~You, W.~Chen, C.~Chen, and Z.~Zheng, ``Who are the
  phishers? phishing scam detection on ethereum via network embedding,''
  \emph{IEEE Transactions on Systems, Man, and Cybernetics: Systems}, vol.~52,
  no.~2, pp. 1156--1166, 2022.

\bibitem{TEGDetector}
J.~Chen, H.~Xiong, D.~Zhang, Z.~Liu, and J.~Wu, ``{T}egdetector: {A} phishing
  detector that knows evolving transaction behaviors,'' \emph{arXiv preprint
  arXiv:2111.15446}, 2021.

\bibitem{li2021self}
S.~Li, F.~Xu, R.~Wang, and S.~Zhong, ``Self-supervised incremental deep graph
  learning for ethereum phishing scam detection,'' \emph{arXiv preprint
  arXiv:2106.10176}, 2021.

\bibitem{kim2018review}
B.~Kim, K.~H. Lee, L.~Xue, and X.~Niu, ``A review of dynamic network models
  with latent variables,'' \emph{Statistics surveys}, vol.~12, p. 105, 2018.

\bibitem{zheng2018blockchain}
Z.~Zheng, S.~Xie, H.-N. Dai, X.~Chen, and H.~Wang, ``Blockchain challenges and
  opportunities: A survey,'' \emph{International journal of web and grid
  services}, vol.~14, no.~4, pp. 352--375, 2018.

\bibitem{chen2020understanding}
T.~Chen, Z.~Li, Y.~Zhu, J.~Chen, X.~Luo, J.~C.-S. Lui, X.~Lin, and X.~Zhang,
  ``Understanding ethereum via graph analysis,'' \emph{ACM Transactions on
  Internet Technology (TOIT)}, vol.~20, no.~2, pp. 1--32, 2020.

\bibitem{holub2018coinhoarder}
A.~Holub and J.~O'Connor, ``Coinhoarder: Tracking a ukrainian bitcoin phishing
  ring dns style,'' in \emph{Proceedings of the 2018 APWG Symposium on
  Electronic Crime Research (eCrime)}.\hskip 1em plus 0.5em minus 0.4em\relax
  IEEE, 2018, pp. 1--5.

\bibitem{gcn}
\BIBentryALTinterwordspacing
T.~N. Kipf and M.~Welling, ``Semi-supervised classification with graph
  convolutional networks,'' in \emph{Proceedings of the 5th International
  Conference on Learning Representations, Conference Track Proceedings}.\hskip
  1em plus 0.5em minus 0.4em\relax OpenReview.net, 2017. [Online]. Available:
  \url{https://openreview.net/forum?id=SJU4ayYgl}
\BIBentrySTDinterwordspacing

\bibitem{vaswani2017attention}
A.~Vaswani, N.~Shazeer, N.~Parmar, J.~Uszkoreit, L.~Jones, A.~N. Gomez,
  {\L}.~Kaiser, and I.~Polosukhin, ``Attention is all you need,'' in
  \emph{Proceedings of the 30th Advances in neural information processing
  systems}, 2017, pp. 5998--6008.

\bibitem{hamilton2017inductive}
W.~Hamilton, Z.~Ying, and J.~Leskovec, ``Inductive representation learning on
  large graphs,'' \emph{Advances in neural information processing systems},
  vol.~30, 2017.

\bibitem{trivedi2019dyrep}
R.~Trivedi, M.~Farajtabar, P.~Biswal, and H.~Zha, ``Dyrep: Learning
  representations over dynamic graphs,'' in \emph{Proceedings of the 7th
  International conference on learning representations}.\hskip 1em plus 0.5em
  minus 0.4em\relax OpenReview.net, 2019.

\bibitem{van2010comparing}
B.~C. Van~Wijk, C.~J. Stam, and A.~Daffertshofer, ``Comparing brain networks of
  different size and connectivity density using graph theory,'' \emph{PloS
  one}, vol.~5, no.~10, p. e13701, 2010.

\bibitem{rossi2020temporal}
E.~Rossi, B.~Chamberlain, F.~Frasca, D.~Eynard, F.~Monti, and M.~Bronstein,
  ``Temporal graph networks for deep learning on dynamic graphs,'' \emph{arXiv
  preprint arXiv:2006.10637}, 2020.

\end{thebibliography}

\end{document}